# Strong Single- and Two-Photon Luminescence Enhancement by Non-Radiative Energy Transfer across Layered Heterostructure

Medha Dandu, Rabindra Biswas, Sarthak Das, Sangeeth Kallatt, Suman Chatterjee, Mehak Mahajan, Varun Raghunathan, Kausik Majumdar*

Department of Electrical Communication Engineering, Indian Institute of Science, Bangalore 560012, India.

*Corresponding author, email: *kausikm@iisc.ac.in*

**ABSTRACT:** The strong light-matter interaction in monolayer transition metal dichalcogenides (TMDs) is promising for nanoscale optoelectronics with their direct band gap nature and the ultra-fast radiative decay of the strongly bound excitons these materials host. However, the impeded amount of light absorption imposed by the ultra-thin nature of the monolayers impairs their viability in photonic applications. Using a layered heterostructure of a monolayer TMD stacked on top of strongly absorbing, non-luminescent, multi-layer $SnSe_2$, we show that both single-photon and two-photon luminescence from the TMD monolayer can be enhanced by a factor of 14 and 7.5, respectively. This is enabled through inter-layer dipole-dipole coupling induced non-radiative Förster resonance energy transfer (FRET) from $SnSe_2$ underneath which acts as a scavenger of the light unabsorbed by the monolayer TMD. The design strategy exploits the near-resonance between the direct energy gap of $SnSe_2$ and the excitonic gap of monolayer TMD, the smallest possible separation between donor and acceptor facilitated by van der Waals heterojunction, and the in-plane orientation of dipoles in these layered materials. The FRET driven uniform single- and two-photon luminescence enhancement over the entire junction area is advantageous over the local enhancement in quantum dot or plasmonic structure integrated 2D layers, and is promising for improving quantum efficiency in imaging, optoelectronic, and photonic applications.

Monolayer transition metal dichalcogenides (TMDs) exhibit direct band gap with reduced dielectric screening and strong quantum confinement that enhance coulomb interactions to form strongly bound excitons under illumination.[1–4] Highly efficient luminescence due to a rapid radiative decay of excitons in TMDs[5–8] instilled extensive interest for using these layered materials in the design of optoelectronic devices. Additionally, TMDs have also been explored for non-linear optical phenomena such as ultrafast saturable absorption, two-photon absorption and harmonic generation which are promising for photonic applications.[9] However, the weak light absorption in a monolayer TMD[10] due to its ultra-thin nature curtails its single and two-photon luminescence regardless of their inherent strength of the light-matter interactions.

Enhanced photoluminescence (PL) from monolayer TMDs has been demonstrated by coupling them with complementing optical platforms like mirco/nano-cavities,[11] photonic crystals[12] and hybrid reflectors.[13,14] Recent integration of plasmonic structures with TMDs have reported very high enhancement factors of PL from TMD.[15,16] This results from a combination of Purcell effect and local enhancement of electric field. Chemically driven surface modification for defect passivation and functionalization with quantum dots are also found to be effective in enhancing TMD PL.[17,18] QDs assembled on monolayer TMDs show PL enhancement in the TMD and PL quenching in the QD coupled with a reduction in its exciton radiative lifetime proving the existence of Förster Resonance Energy Transfer (FRET).[19] Such non-radiative energy transfer is enabled by dipole-dipole coupling between the donor and the acceptor and its rate is governed by their spectral overlap, physical separation, orientation of dipoles and corresponding oscillator strengths.[20]

It would be interesting to exploit FRET across layered heterostructures where each layer has large excitonic oscillator strengths with a possibility of matching of in-plane center of mass momentum, facilitating dipolar coupling.[21] Kozawa *et al.*[22] showed the evidence for fast interlayer energy transfer across $WS_2/MoSe_2$ hetero-bilayer stack through FRET across higher order exciton transitions. Nonetheless, the donor's absorption is still constrained by its physical thickness at the monolayer limit despite the high efficiency of FRET at the closest possible physical separation.

Here, we demonstrate enhanced PL of monolayer $MoS_2$ (and $WS_2$) across its vdW heterojunction with multi-layer $SnSe_2$ via FRET with single and two-photon excitation. Counteracting the charge transfer across highly staggered conduction bands of $MoS_2$ and $SnSe_2$, $MoS_2$ single-photon luminescence (1P-PL) shows ~14-fold enhancement at room temperature with resonant excitation and ~5-fold enhancement with non-resonant excitation, while two-photon luminescence (2P-PL) of $MoS_2$ shows up to ~7.5-fold enhancement with non-resonant excitation. Even with the insertion of few-layer hBN between $MoS_2$ and $SnSe_2$, the 1P-PL enhancement persists up to 5 times with resonant excitation. We demonstrate modulation of the degree of the PL enhancement by systematic parameter variation, including donor material, acceptor material, their thickness, physical separation between donor and acceptor, sample temperature, and excitation wavelength which corroborate FRET aided PL enhancement. We emphasize the intrinsic advantage of realizing FRET with $SnSe_2$ as a donor and elucidate the impact of multiple parameters on the luminescence enhancement using rate equation analysis.

**Results and Discussions:**

FRET allows transitions in the acceptor resonant to the energy of the excited states in the donor. In the approximation of weak coupling, the rate of FRET depends on the spectral overlap between the donor's emission and the acceptor's absorption, the relative orientation of the dipoles across the donor and the acceptor and the physical separation between them.[23,24] Considering the above factors, we realize FRET across a vdW heterostructure with monolayer $MoS_2$ (1L-$MoS_2$) or $WS_2$ (1L-$WS_2$) as acceptor, stacked on top of multi-layer $SnSe_2$ as donor, which acts as a scavenger of the light unabsorbed by the top monolayer. With vdW stacks of layered materials, FRET benefits from the in-plane orientation of dipoles in both the donor and the acceptor coupled with the closest possible vertical spacing, as depicted in Figure 1a. While $SnSe_2$ is an indirect bandgap material with an energy gap of 1.1 eV, the choice of $SnSe_2$ stems from its direct gap at the 'K' point[25–30] which is close to the excitonic gap of 1L-$MoS_2$[31] and 1L-$WS_2$,[32] as shown in Figure 1b. $SnSe_2$ also exhibits nearly flat bands at the direct gap facilitating strong absorption of light. Due to its indirect nature, we did not obtain any luminescence from $SnSe_2$, down to 4 K.

**Single photon luminescence enhancement:** We first perform 1P-PL measurement on the isolated 1L-$MoS_2$ (denoted as M) and the 1L-$MoS_2$/$SnSe_2$ junction (denoted as MS) at room temperature (RT) with resonant (633 nm) and non-resonant (532 nm) single-photon excitations. Illumination of MS region with 532 nm laser, which is non-resonant to both $MoS_2$ and $SnSe_2$, creates carriers much above their direct transition, while 633 nm near-resonantly excites both $MoS_2$ and $SnSe_2$. Figure 1c shows the PL map (embedded in the optical image) of the MS stack (sample 1, denoted as J1) under 633 nm excitation (See Figure S1a in **Supporting Information** for the optical image). The green dotted line in the image depicts the periphery of 1L-$MoS_2$ transferred on top of $SnSe_2$ (orange dotted boundary) using dry transfer method (See **Methods** for heterojunction fabrication and Figure S1b in **Supporting Information** for Raman characterization). PL mapped area covers individual 1L-$MoS_2$, $SnSe_2$ and junction regions, and shows strong and uniform enhancement in the junction region. Figure 1d shows a representative PL spectrum of isolated $MoS_2$ (M-J1) and junction (MS-J1) regions corresponding to the PL map in Figure 1c. Fitting of the above spectra for individual exciton and trion peaks[17,33] gives about 14 times enhancement for $A_{1s}$ exciton peak of MS-J1. When the excitation wavelength is switched to 532 nm, the PL enhancement persists as shown in Figure 1e but the enhancement of $A_{1s}$ peak goes down to a factor of ~5. On the contrary, trion (T) peak intensity remains nearly unchanged and $B_{1s}$ peak gets quenched. The strong PL enhancement of the MS region is verified from two other samples, J2 and J3 (See Figure S2 in **Supporting Information**).

Interestingly, the MS junction forms a type-I heterojunction with ~1 eV CB offset, as shown in Figure 1b. The band offsets force fast charge transfer from $MoS_2$ to $SnSe_2$,[34] which is expected to quench the $MoS_2$ PL strength significantly, instead of enhancing. Such charge transfer driven PL quenching has been widely reported in the past.[35–37] The observation of strong PL enhancement despite such charge transfer suggests that $SnSe_2$ is playing a crucial role in the enhancement mechanism. In order to verify this, we replace $SnSe_2$ by 1T-$TaS_2$,[38,39] a material that exhibits relatively weak light absorption. Figure 1f depicts the PL spectra of isolated 1L-$MoS_2$ (M-J4) and 1L-$MoS_2$/$TaS_2$ (MT-J4) with 633 nm excitation (See Figure S1c in **Supporting Information** for an optical image of junction J4) – clearly indicating quenching of PL in the junction. Also, we observe PL quenching in 1L-$MoS_2$/graphene, 1L-$WS_2$/graphene, and 1L-$MoS_2$/$In_{0.53}Ga_{0.47}As$ junctions compared to corresponding isolated monolayers, further indicating the crucial role played by $SnSe_2$ in the observed PL enhancement.

**Evidence of FRET mechanism:** In order to rule out any inter-layer charge transfer and validate the FRET driven non-radiative energy transfer as the dominant PL enhancement mechanism, a ~10 nm thick hBN film is inserted as a barrier layer between 1L-$MoS_2$ and $SnSe_2$, which blocks any possible inter-layer charge transfer. Figure 2a shows the 1L-$MoS_2$/hBN/$SnSe_2$ junction (J5) with highlighted boundaries of each layer. The corresponding PL mapping taken across the portions of 1L-$MoS_2$/hBN (MH-J5) and 1L-$MoS_2$/hBN/$SnSe_2$ (MHS-J5) with 633 nm excitation suggests that $MoS_2$ PL enhances strongly on the latter portion with respect to 1L-$MoS_2$/hBN. The thickness of the hBN and the $SnSe_2$ layers can be inferred from the AFM scans along dotted lines

A and B, as shown in the inset. Figure 2b depicts the representative spectra of the different regions of varying $SnSe_2$ thickness, along the dotted arrow B in Figure 2a. Similar PL enhancement is demonstrated across the MHS region in another sample J6 shown in Figure S3 of **Supporting Information**. These observations corroborate that the PL enhancement is driven by near-field dipolar coupling across 1L-$MoS_2$/$SnSe_2$ that exists even in the presence of an energy barrier like hBN.

The second, third and fourth panels of Figure 2b show a systematic reduction of the PL intensity with an increase in the $SnSe_2$ thickness. While heterojunctions with thin $SnSe_2$ show PL enhancement, with thicker $SnSe_2$ (beyond ~20 nm), we observe a consistent quenching of the PL at the junction compared to isolated $MoS_2$. A similar trend is maintained even when the hBN spacer was removed (see Table 1). Such a systematic variation of PL intensity with the $SnSe_2$ film thickness provides another direct evidence of FRET as explained below.

Figure 2c shows reflectance contrast spectra of isolated $SnSe_2$ films with varying thickness placed on $SiO_2$/Si substrate (see **Methods** for measurement details), measured at room temperature. The dip in the reflectance contrast is found to be efficiently modulated by the thickness of $SnSe_2$. Note that this dip is spectrally broad and does not show any sharp excitonic feature (even down to 4 K, see Figure S4 in **Supporting Information**). This is in agreement with our previous KPFM measurements which show that the $SnSe_2$ is degenerately n-doped with the Fermi level about 0.3 eV above conduction band (CB) minimum.[40] The screening offered by the large carrier concentration suppresses strongly bound exciton state in $SnSe_2$. Throughout this paper, we thus assume that the transition dipole in $SnSe_2$ is governed by free electron-hole pairs.

The absorption peaks corresponding to the minima in the reflectance contrast spectra are plotted in Figure 2d as a function of $SnSe_2$ thickness. In the same plot, we also show the $A_{1s}$ excitonic absorption peaks of monolayers of $MoS_2$, $WS_2$ and $WSe_2$, at room temperature. Along with the spectral overlap between the donor and the acceptor, strength of absorption in the donor is an important parameter that controls the efficiency of energy transfer through FRET process. The diagram suggests that by changing the thickness of $SnSe_2$ film, one can effectively tune the strength of absorption at corresponding direct gap in $SnSe_2$ (donor) and the spectral overlap between $SnSe_2$ (donor) and the monolayer acceptor, in turn controlling the non-radiative energy transfer efficiency. In particular, as the thickness of $SnSe_2$ film is increased beyond ~20 nm, the FRET efficiency is reduced between $SnSe_2$ direct gap absorption and the $MoS_2$ $A_{1s}$ excitonic absorption peak, and is in excellent agreement with our PL data.

In agreement with the above argument, replacing 1L-$MoS_2$ by 1L-$WS_2$ (with a 2.02 eV excitonic gap at RT) yields a strong PL enhancement as well. This is shown by the PL mapping image of the $WS_2$/$SnSe_2$ stack J7 in Figure 2e (See Figure S1d in **Supporting Information** for its optical image). We obtain ~3 times enhancement (with off-resonant 532 nm excitation) of $WS_2$ PL on the junction with thin $SnSe_2$ (region A) and the enhancement goes down when $SnSe_2$ becomes thicker (region B and C). A representative spectrum is shown in the middle panel of Figure 2f. In contrast, when the resonance between monolayer TMD and $SnSe_2$ is avoided by using 1L-$WSe_2$/$SnSe_2$ junction J8, where excitonic gap of $WSe_2$ (~1.65 eV at RT) is much below the $SnSe_2$ direct gap,[32] no PL enhancement is observed, as seen from the right most panel of Figure 2f.

Further, we observe a strong temperature dependence of the PL enhancement factor on the MS junction region. Figure 3a and 3b show the corresponding spectra of M-J2 and MS-J2 under 633 nm and 532 nm excitations, up to 200 K. In both the cases, the ratio ($\alpha$) of the peak intensity on the junction to that on the 1L-$MoS_2$ of both exciton ($A_{1s}$) and trion increase with temperature. At low temperatures below 100K, $MoS_2$ PL intensity on the junction is reduced drastically for both $A_{1s}$ and trion peaks. Above 100K, while the trion peak remains quenched on the junction, the $A_{1s}$ peak sharply increases with temperature. Figure 3c depicts the temperature variation of α for $A_{1s}$ and trion of 1L-$MoS_2$/$SnSe_2$ (MS-J2) under 532 nm excitation. Similar temperature dependence of α for $A_{1s}$ is also seen on 2L-$MoS_2$/$SnSe_2$ junction where the degree of quenching increases with a reduction in the temperature (See Figure S5 in **Supporting Information**). As explained in Figure S6b of **Supporting Information**, **s**uch a strong dependence on temperature[41] can be directly correlated with a temperature dependent rate of the non-radiative inter-layer energy transfer. Figure 3c also illustrates the variation of α with temperature for $A_{1s}$ on 1L-$MoS_2$/$TaS_2$ (MT-J4) under 633 nm excitation, which, on the contrary, does not show any temperature dependence of the degree of PL quenching, unlike $MoS_2$/$SnSe_2$ junction.

**Other possible mechanisms as play:** Constructive interference with the back-reflected light from the substrate is one mechanism often used to enhance the PL strength. Two experimental observations also help us to rule out the contribution of interference effect in PL enhancement.. First, as mentioned above, the PL enhancement at the $MoS_2$/$SnSe_2$ junction shows a strong temperature dependence. Any such constructive interference with back reflected light is not expected to be a strong function of sample temperature. Second, we observe a strong quenching of the $MoS_2$ Raman peaks at the $MoS_2$/$SnSe_2$ junction (See Figure S1b in **Supporting Information**) while the PL intensity is enhanced. Any optical effect like constructive interference would be similar for both these cases, indicating a different origin for PL enhancement. Raman peaks are strongly quenched since the carrier scattering with optical phonons is a relatively slower process compared to ultra-fast inter-layer charge and energy transfer.

Another possible mechanism could be enhanced quantum yield at the heterojunction due to isolation from $SiO_2$ trap induced inhomogeneity. While such inhomogeneities can play an important role at low temperature, exciton-phonon scattering plays a dominant role at room temperature.[42] The observation of strong PL enhancement at room temperature thus indicates a relatively small effect of changes in quantum yield. In addition, any such enhanced quantum yield cannot explain the $SnSe_2$ thickness dependence of the PL strength at the heterojunction in Figure 2a-b, since in all these cases, 1L-$MoS_2$ is sitting on top of hBN/$SnSe_2$ stack, hence the quantum yield is expected to be similar.

**Modeling the processes using rate equations:** In Table 1, we summarize the experimental observations over different stacks, and dependence of PL strength on excitation wavelength (resonant versus non-resonant), sample temperature, donor and acceptor materials, thickness of the donor and the acceptor, and the separation between them. We now derive an expression for the PL enhancement factor (α) from a rate equation analysis which provides an insight into the effect of various parameters on α. The different processes occurring across $MoS_2$, $SnSe_2$ and the junction on illumination are represented in the transition picture with resonant excitation in Figure 4a. $\Gamma_r$ and $\Gamma_{nr}$ are respectively the radiative and non-radiative decay rates of exciton in $MoS_2$, while $\Gamma_s$ is the scattering rate of transition dipoles in $SnSe_2$. Let $\Gamma_{CT}^{M-S}$ and $\Gamma_{ET}^{M-S}$ denote the rate of charge and energy transfer respectively from $MoS_2$ to $SnSe_2$. The corresponding energy transfer rate from $SnSe_2$ to $MoS_2$ is denoted by $\Gamma_{ET}^{S-M}$. The detailed analysis of the rate equations in $MoS_2$ and the junction considering all the relevant processes as indicated in Figure 4a is presented in **Supporting Information**. α is deduced as

$$\alpha = \left(\frac{G_{e-h}^{S}}{G_{ex}^{M}}\right)\frac{\gamma}{\gamma + \beta} \tag{1}$$

where $\gamma = \frac{\Gamma_{ET}^{S-M}}{\Gamma_s}$ , $\beta = 1 + \left(\frac{\Gamma_{CT}^{M-S}+\Gamma_{ET}^{M-S}}{\Gamma_r+\Gamma_{nr}^{M}}\right)$, $G_{e-h}^{S}$ is the effective rate of generation of transition dipoles in $SnSe_2$ at the energy states resonant to $A_{1s}$ excitonic states of $MoS_2$ and $G_{ex}^{M}$ is the effective rate of generation of excitons in $MoS_2$. γ represents the ratio of the rate of resonant energy transfer from $SnSe_2$ to $MoS_2$, to the relaxation rate to the indirect band within $SnSe_2$. A large value of γ is desirable, which is achieved by the ultra-fast inter-layer energy transfer rates in vdW heterostructures as theoretically predicted in literature.[22,36,43,44] Clearly, α takes a maximum value of $\frac{G_{e-h}^{S}}{G_{ex}^{M}}$ when $\beta \ll \gamma$.

Resonant and non-resonant excitations of 1L-$MoS_2$/$SnSe_2$ differ in the factor $\frac{G_{e-h}^{S}}{G_{ex}^{M}}$ that changes α in (1). As non-resonant excitation creates e-h pairs in $MoS_2$ and $SnSe_2$ at much higher energies above the direct transitions, scattering of these hot carriers lowers both $G_{e-h}^{S}$ and $G_{ex}^{M}$. Also, as few-layer hBN is inserted between 1L-$MoS_2$ and $SnSe_2$, $\Gamma_{CT}^{M-S}$ is completely suppressed, however, coupled with a reduction in $\Gamma_{ET}^{M-S}$ due to increased separation ($d$).[41] This effectively results in a lower value of α at 1L-$MoS_2$/hBN/$SnSe_2$ junction than that of 1L-$MoS_2$/$SnSe_2$ junction.

The dissimilarity in the degree of enhancement for $A_{1s}$ exciton and trion can be explained with the competition between $\Gamma_r$ and the rate of FRET as elucidated in Figure S6c of **Supporting Information**. As $A_{1s}$ exciton exhibits faster radiative decay ($\Gamma_r$), PL enhancement persists despite the intrinsic charge ($\Gamma_{CT}^{M-S}$) and energy ($\Gamma_{CT}^{M-S}$) transfer from $MoS_2$ to $SnSe_2$. On the other hand, trions possess slower $\Gamma_r$ [8] and hence suffer from $\Gamma_{CT}^{M-S}$ and $\Gamma_{ET}^{M-S}$ because of their longer lifetime. This effectively increases β in (3) resulting in suppressed or no enhancement for the trion. This points to the significance of the oscillator strength of transition dipoles in the acceptor for efficient FRET induced PL enhancement.

**Efficient FRET due to additional degree of freedom in $SnSe_2$:** From a microscopic point of view, energy and center of mass (COM) momentum must be conserved between the donor and the acceptor dipoles during the FRET process.[41] The layered heterostructure forces the dipoles to be oriented in the plane, facilitating the conservation of the in-plane momentum, as illustrated in Figure 1a. Since there is a mismatch ($\Delta E_g$) between the direct bandgap in $SnSe_2$ and exciton gap in 1L-$MoS_2$, coupled with a mismatch in the effective mass values in these materials, it is difficult to conserve both energy and momentum simultaneously when it comes to exciton-exciton energy transfer, as pointed out by Lyo.[41] To this end, the transition dipoles in $SnSe_2$ being governed by e-h pair as explained earlier (and not strongly bound exciton due to screening), allows an additional degree of freedom in the form of the relative motion between the electron and the hole.

The situation is schematically illustrated in Figure 4b. The violet color band on the left is the exciton band structure of $MoS_2$, with the energy given by $E = \frac{\hbar^2 K_m^2}{2M_m}$, where $\mathbf{K_m}$ is the COM momentum of the $MoS_2$ exciton, and $M_m = m_{e,MoS_2} + m_{h,MoS_2}$. In $SnSe_2$, the total energy of the e-h pair is given by the sum of the COM component (in orange) and the relative component (in blue): $E = \frac{\hbar^2 K_s^2}{2M_s} + \frac{\hbar^2 k_s^2}{2\mu_s}$ where $\mathbf{K_s}$ is the in-plane COM momentum, $\mathbf{k_s}$ is the relative part of the in-plane momentum, and $\mu_s = \frac{m_{e,SnSe_2} m_{h,SnSe_2}}{M_s}$ is the reduced mass, with $M_s = m_{e,SnSe_2} + m_{h,SnSe_2}$. Energy conservation during FRET at a COM momentum $\mathbf{K_m} = \mathbf{K_s} = \mathbf{K_0}$ yields

$$\frac{\hbar^2 K_0^2}{2M_m} = \Delta E_g + \frac{\hbar^2 K_0^2}{2M_s} + \frac{\hbar^2 k_s^2}{2\mu_s} \tag{2}$$

Equation (2) clearly shows that the free variable $k_s$ due to the relative movement between electron and hole in $SnSe_2$ relaxes the simultaneous energy-momentum conservation condition, improving the FRET efficiency. For a given $K_0$, we have

$$k_s = \frac{2\mu_s}{\hbar^2}\sqrt{\frac{\hbar^2 K_0^2}{2}\left(\frac{1}{M_m} - \frac{1}{M_s}\right) - \Delta E_g} \tag{3}$$

where solutions exist for $k_s$ if $\left[\frac{\hbar^2 K_0^2}{2}\left(\frac{1}{M_m} - \frac{1}{M_s}\right) - \Delta E_g\right] > 0$. In the present case, we have $M_m < M_s$, and $\Delta E_g > 0$. Also, note that equation (2) only has a constraint on the absolute value $k_s$, hence all possible angular directions in the plane of the layer can contribute to FRET, as illustrated in Figure 4b.

**PL enhancement with two-photon excitation:** We now show the generic nature of the proposed method by demonstrating enhanced luminescence using two-photon absorption (TPA). TPA is a third order non-linear process where excitation is performed with two photons of longer wavelength compared to the bandgap. TPA in materials with optical band gap in the visible range is attractive for dual mode visible-IR photodetection.[45] TPA is useful to access the energy states such as dark excitons which are forbidden by selection rules under single-photon excitation.[46,47] Two-photon luminescence (2P-PL) is also widely sought in biological sensing and high-resolution imaging due to axial localization and large penetration depth.[48] The FRET process across layered heterostructures is a very versatile method of energy transfer that is applicable to two-photon absorption as well. This allows facile integration of a material with strong TPA as a donor layer with a highly luminescent medium like TMD monolayer, as an acceptor, and is promising for such imaging and spectroscopic applications.[49,50] Such mechanism can also yield enhanced frequency up-conversion. Recent reports have shown utilization of donors with higher two-photon absorption cross-section to yield higher 2P-PL.[51,52]

We explore this phenomenon across 1L-$MoS_2$/$SnSe_2$ (sample J1) by two-photon PL imaging using 1040 nm excitation (See **Methods**). We observe a 7.5-fold 2P-PL enhancement as shown in the 2P-PL map of J1 in Figure 5a where the emitted light is filtered at 650 nm with a band pass window of 40 nm. Figure 5b shows the voltage signal of the PMT that detects 2P-PL from the isolated 1L-$MoS_2$ (violet) and the junction (orange) as a function of peak irradiance. The enhancement is particularly strong at lower peak irradiance, which is generally desirable to obtain strong 2P-PL without possible sample damage. The 2P-PL signal from the isolated 1L-$MoS_2$ portion shows non-linear variation with peak irradiance. On the other hand, 2P-PL from the heterojunction, while showing strong enhancement compared to the isolated $MoS_2$ portion, does not exhibit any such non-linearity in the measured range. This suggests that 1L-$MoS_2$ on the junction gets excited by the cascaded process of TPA in $SnSe_2$ followed by carrier relaxation and subsequent FRET from $SnSe_2$ to $MoS_2$, as schematically depicted in Figure 5c. Such 2P-PL enhancement is also observed across the 1L-$MoS_2$/hBN/$SnSe_2$ junction. Thus, FRET with two-photon excitation can aid to increase the saturation threshold of TPA in TMD monolayer besides enhancing 2P-PL significantly at lower power levels. This way of enhancing the luminescence with multi-photon excitation across layered heterostructures is interesting for nanoscale photonic applications.[9]

**Conclusion:**

In summary, using a layered heterostructure of monolayer TMD and multi-layer $SnSe_2$, we demonstrated a strongly enhanced luminescence manifested through non-radiative resonance energy transfer. This mechanism benefits from the following factors: 1) nanometer scale separation between the donor and the acceptor layer; 2) in-plane orientation of dipoles, both in the donor and the acceptor, providing improved dipolar coupling; 3) scavenging of light that is unabsorbed by monolayer TMD using multi-layer $SnSe_2$ which also has a near-resonant direct transition to exciton state of the corresponding TMD; 4) energy transfer from free e-h pairs in $SnSe_2$ to strongly bound excitons in monolayer TMD relaxing energy conservation due to additional degree of freedom. The generic nature of the mechanism is attractive, and is demonstrated for both single- and two-photon luminescence enhancement. The luminescence enhancement being uniform across the entire overlapped junction area is advantageous over the local enhancement in plasmonic structure or QD coupled 2D layers. The general strategy of combining a weak absorber, but highly luminescent material (like monolayer TMD) with a strong optical absorber devoid of luminescence (like $SnSe_2$) can be extended to other material systems as well, and holds promise for improved quantum efficiency in multiple applications such as biomedical imaging, light emitting diode, light harvesting, and non-linear spectroscopy. Also, in order to further improve the overall photoluminescence from monolayer TMDs, the scheme presented here can work in tandem with other techniques (for example, defect passivation[18]) that help to improve the quantum yield in the monolayer.

**Methods:**

**Fabrication of heterojunction:** $MoS_2/SnSe_2$ stack is prepared by all-dry transfer method using visco-elastic stamping maneuvered by micromanipulator. First, $SnSe_2$ is exfoliated on to 285 nm $SiO_2$ coated Si substrate. $MoS_2$ is exfoliated on PDMS attached to a glass slide and the $MoS_2$ flake of desired thickness is identified and then transferred on top of multi-layer $SnSe_2$ flake of interest. After transfer, $MoS_2/SnSe_2$ stack is measured without any annealing. However, remeasurements after heating the stack did not differ significantly with the measurements of pre-heated stack.

**Reflectance contrast measurement:** For reflectance contrast measurements, $SnSe_2$ flakes of varying thickness are exfoliated on top of 285 nm $SiO_2$ coated Si substrate. Thickness of $SnSe_2$ flakes is determined with Atomic Force Microscopy. A broadband radiation is passed through a pinhole and focused on to the $SnSe_2$ flake of interest using 100x objective. The reflected light is collected in a confocal mode using a spectrometer with 1800 lines per mm grating and CCD. Reflectance spectrum from the bare $SiO_2$ substrate ($R_{SiO_2}$) is subtracted from the reflectance spectrum of individual $SnSe_2$ flake ($R_{Sample}$) and normalized with $R_{SiO_2}$ to identify the absorption energy position of $SnSe_2$. Reflectance contrast measurements at 4K are taken in a similar way as above using a 50x objective to focus the light on to the sample in a closed-cycle cryostat.

**Two photon PL microscopy:** The two photon PL imaging of the hetero junction is performed using a fiber laser source (Coherent Fidelity HP-10) operating at 1040 nm wavelength with 80 MHz repetition rate. The horizontally polarized laser of 140 fs pulse width is focused onto the sample using a ~20x/0.75 objective (Olympus UPLSAPO 20x). The two-photon PL signal is collected using the same objective and then separated from the excitation beam using a dichroic mirror and detected using a photomultiplier tube (Hamamatsu R3896). A 650 nm/40 nm band-pass filter in combination with an 890 nm short-pass filter is mounted in front of the photomultiplier tube for efficient rejection of any residual excitation beam and to minimize the background signal. Two-photon PL images are acquired by scanning a pair of galvanometric mirrors (Thorlabs GVS002) over an imaging area of 15 µm x15 µm.

**Acknowledgements:**

K. M. acknowledges the support a grant from Indian Space Research Organization (ISRO), grants under Ramanujan Fellowship, Early Career Award, and Nano Mission from the Department of Science and Technology (DST), Government of India, and support from MHRD, MeitY and DST Nano Mission through NNetRA.

**Conflicts of Interest:**

The authors declare no financial or non-financial conflict of interest.

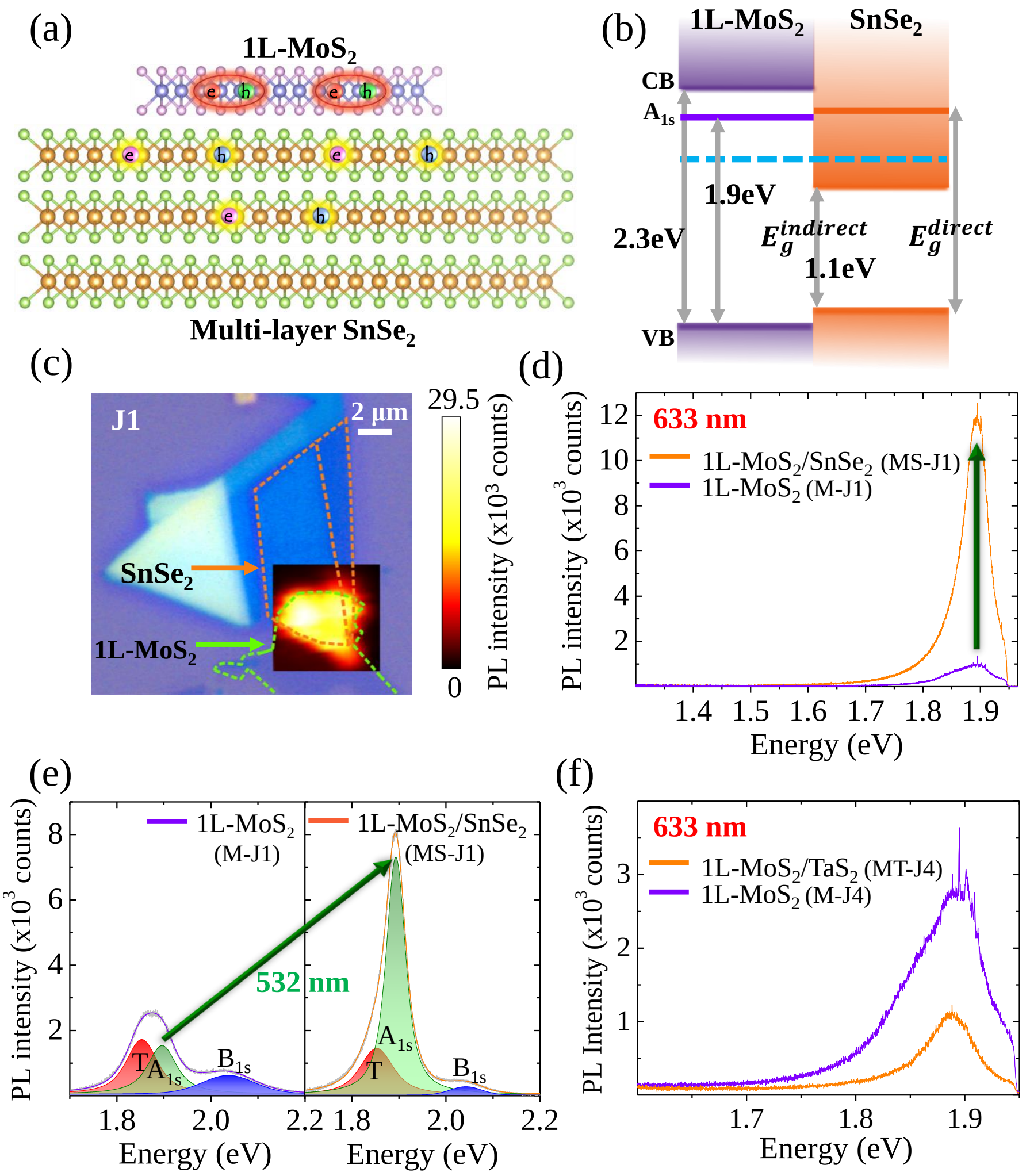

**Figure 1: Single-photon luminescence enhancement of 1L-$MoS_2$ on $SnSe_2$ at room temperature**. (a) In-plane orientation of dipoles across layered heterostructure which can maximize the efficiency of FRET. (b) Band diagram of 1L-$MoS_2$/$SnSe_2$ junction showing type-I heterojunction. In $MoS_2$, the 1s state of the A exciton ($A_{1s}$) and the continuum are around 1.9 eV and 2.3 eV. The direct and indirect energy gaps in $SnSe_2$ are also shown. (c) Optical image of the 1L-$MoS_2$/$SnSe_2$ junction (sample J1) with embedded PL map at 1.89 eV under 633 nm excitation at room temperature. Regions highlighted with orange and green dotted lines represent $SnSe_2$ and 1L-$MoS_2$, respectively. Color bar maps the PL intensity counts. (d) PL spectra from isolated 1L-$MoS_2$ on $SiO_2$ (violet) and 1L-$MoS_2$/$SnSe_2$ (orange) of sample J1 under illumination with 633 nm laser. (e) PL spectra from isolated 1L-$MoS_2$ on $SiO_2$ (left panel) and 1L-$MoS_2$/$SnSe_2$ (right panel) of sample J1 under 532 nm excitation shown along with fitting of corresponding trion (T) and 1s exciton ($A_{1s}$ and $B_{1s}$) peaks. Grey shaded line represents raw spectral data. (f) PL spectra obtained with 633 nm excitation from isolated 1L-$MoS_2$ on $SiO_2$ (violet) and 1L-$MoS_2$/$TaS_2$ junction (orange) of sample J4.

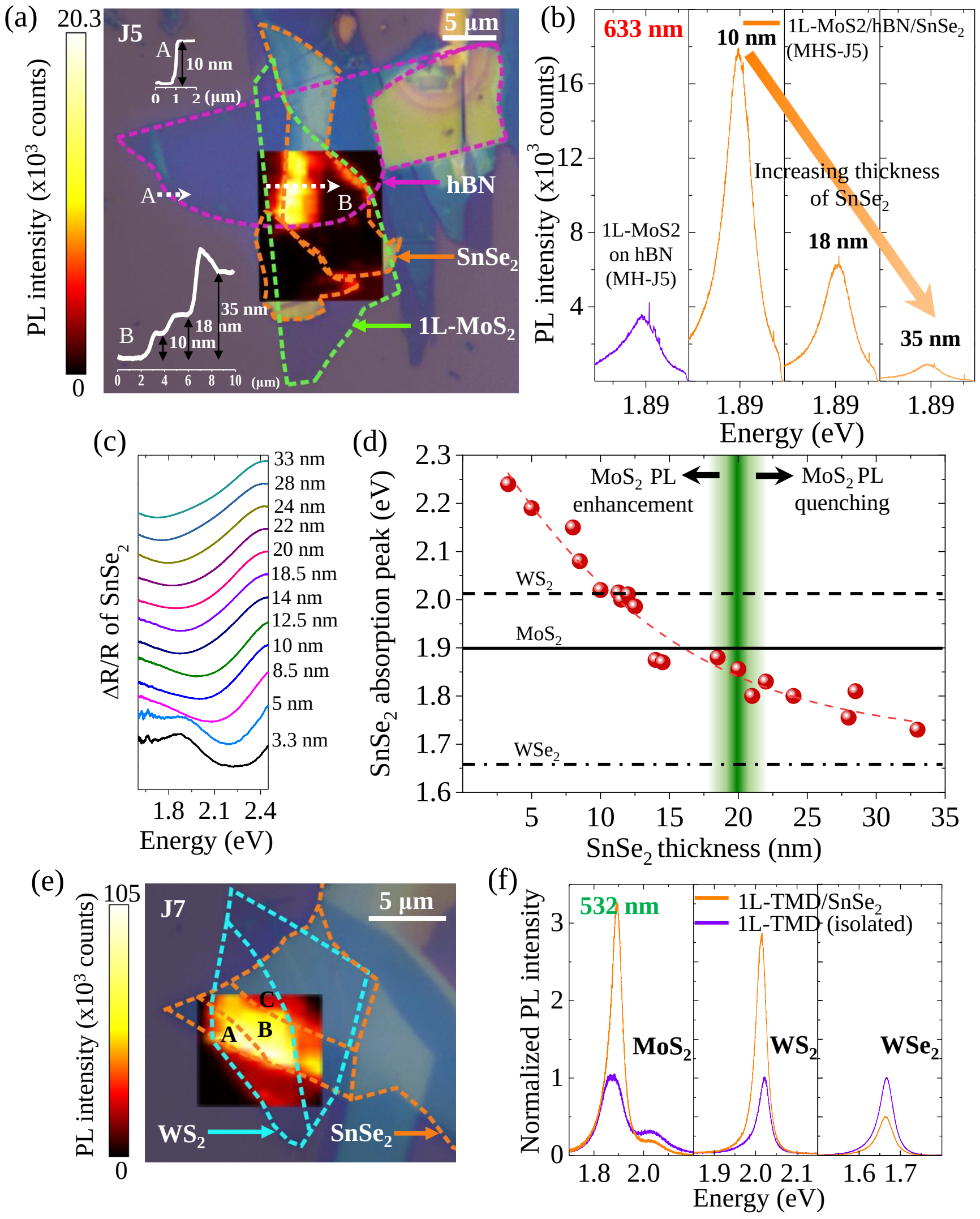

**Figure 2: Evidence of FRET mechanism**. (a) 1L-$MoS_2$/hBN/$SnSe_2$ junction (sample J5) with its optical image and PL map under 633 nm illumination. 1L-$MoS_2$ (green), hBN (magenta) and $SnSe_2$ (orange) layers are individually highlighted in the optical image. Thickness of hBN (top inset) and $SnSe_2$ (bottom inset) layers determined from the step heights in AFM scans along the lines A and B, respectively. (b) PL spectra of 1L-$MoS_2$/hBN (in violet, left column) and 1L-$MoS_2$/hBN/$SnSe_2$ (in orange) with different thickness values of $SnSe_2$ from sample J5 [along the white dashed arrow B in (a)]. PL intensity of 1L-$MoS_2$ on the junction decreases with increase in $SnSe_2$ thickness. (c) Reflectance contrast $\frac{\Delta R}{R} = \frac{R_{SnSe_2} - R_{SiO_2}}{R_{SiO_2}}$ of isolated $SnSe_2$ on $SiO_2$ at room temperature. Corresponding thickness of $SnSe_2$ is indicated adjacent to each spectrum. (d) Absorption peak of $SnSe_2$ extracted from reflectance contrast as a function of its thickness. The solid, dashed and dot-dashed lines highlight the excitonic resonances of $MoS_2$, $WS_2$ and $WSe_2$ at room temperature. The green gradient line partitions the $SnSe_2$ thickness regions for $MoS_2$ PL enhancement and quenching as observed experimentally. (e) Optical image of the 1L-$WS_2$/$SnSe_2$ junction (sample J7) with embedded PL map at 2.02 eV under 532 nm excitation at room temperature. Orange and blue dotted lines highlight the portions of $SnSe_2$ and $WS_2$, respectively. Region A, B and C represent the $WS_2$/$SnSe_2$ regions with increasing $SnSe_2$ thickness. (f) PL spectra from isolated 1L-TMD on $SiO_2$ (violet) and 1L-TMD/$SnSe_2$ junction (orange) of samples J1 ($MoS_2$/$SnSe_2$), J7 ($WS_2$/$SnSe_2$) and J8 ($WSe_2$/$SnSe_2$) with 532 nm excitation.

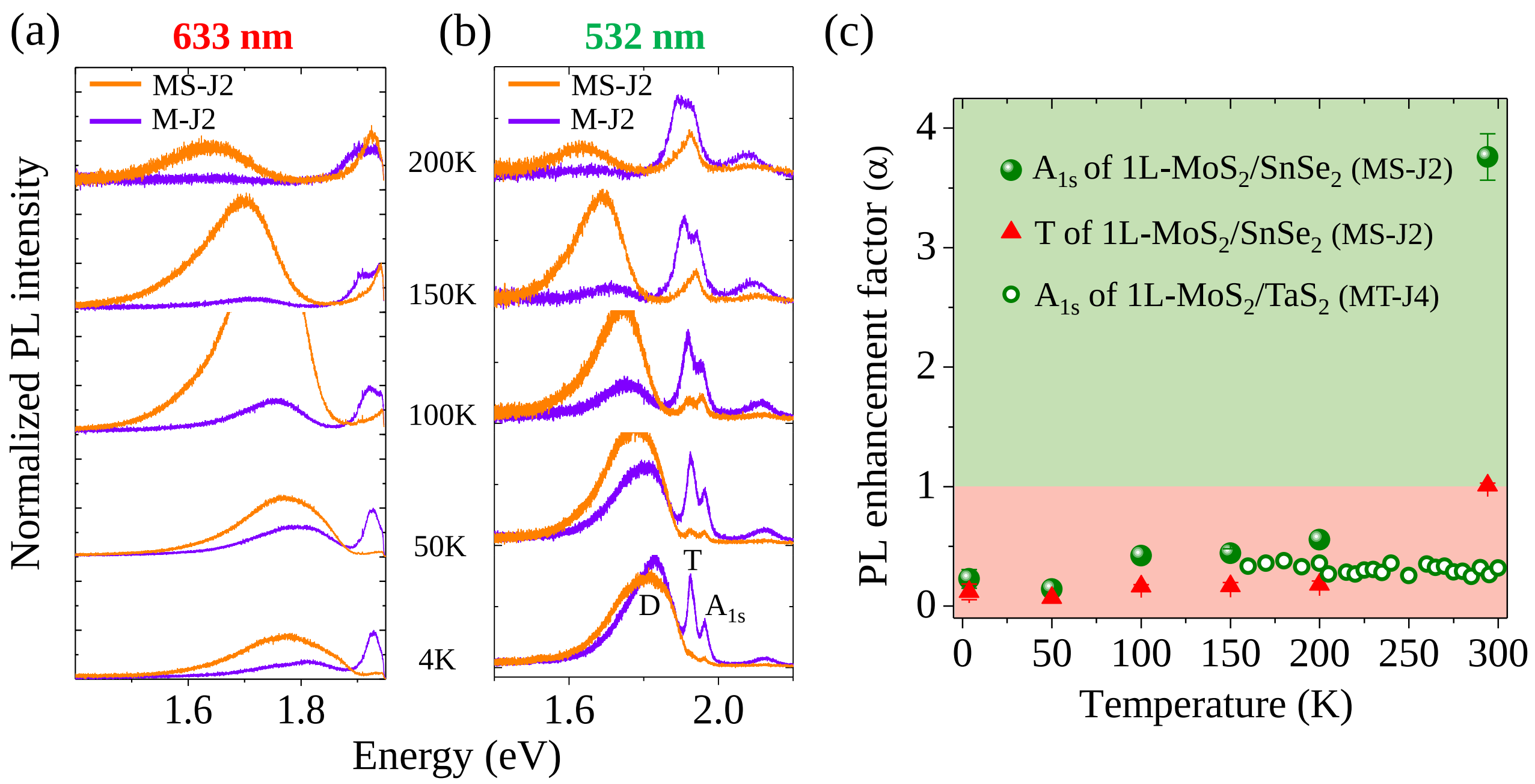

**Figure 3: Temperature dependence of PL enhancement**. (a-b) PL spectra from isolated 1L-$MoS_2$ on $SiO_2$ (violet) and 1L-$MoS_2$/$SnSe_2$ (orange) of sample J2 at multiple temperatures with (a) 633 nm and (b) 532 nm excitation respectively. PL spectra are normalized with respect to the maximum PL intensity from isolated 1L-$MoS_2$ at every temperature. The defect (D), trion (T) and neutral A exciton ($A_{1s}$) are indicated in (b). (c) Temperature dependence of PL enhancement factor ($\alpha$) of $A_{1s}$ exciton and trion peaks of 1L-$MoS_2$/$SnSe_2$ junction (J2) with 532 nm excitation, and $A_{1s}$ exciton peak of 1L-$MoS_2$/$TaS_2$ (J4) with 633 nm excitation.

| | Stack (A/B) | PL enhancement factor (α) | | | | | | | |
|---|---|---|---|---|---|---|---|---|---|
| | | Resonant excitation (633 nm) | | | | Non-resonant excitation [532 nm / (1040/2) nm] | | | |
| | | 150K | | 294K | | 150K | | 294K | |
| | | $t_B$ ~ 8-20 nm | $t_B$ ~ 20-40 nm | $t_B$ ~ 8-20 nm | $t_B$ ~ 20-40 nm | $t_B$ ~ 8-20 nm | $t_B$ ~ 20-40 nm | $t_B$ ~ 8-20 nm | $t_B$ ~ 20-40 nm |
| Single - photon excitation | 1L-$MoS_2$/$SnSe_2$ | 0.9 | --- | 14 | 0.75 | 0.42 | --- | 5 | 0.35 |
| | 1L-$WS_2$/$SnSe_2$ | --- | --- | --- | --- | 0.64 | --- | 3 | 0.62 |
| | 1L-$WSe_2$/$SnSe_2$ | --- | --- | 0.96 | --- | ---- | --- | 0.5 | --- |
| | 2L-$MoS_2$/$SnSe_2$ | --- | 0.23 | --- | 0.45 | --- | 0.17 | --- | 0.3 |
| | 1L-$MoS_2$/hBN/$SnSe_2$ | --- | --- | 4 | 0.182 | --- | --- | 0.54 | 0.154 |
| | 1L-$MoS_2$/$TaS_2$ | --- | 0.35 | --- | 0.32 | --- | 0.32 | --- | 0.33 |
| Two - photon excitation | 1L-$MoS_2$/$SnSe_2$ | --- | --- | --- | --- | --- | --- | 7.5 | --- |
| | 1L-$MoS_2$/hBN/$SnSe_2$ | --- | --- | --- | --- | --- | --- | 2.4 | --- |

**Table 1: Luminescence enhancement summary**. Summary of averaged luminescence intensity enhancement factor ($\alpha$) values from multiple layered heterostructures measured under varying excitation types (single versus two photons), excitation wavelength (resonant versus non-resonant), temperature, and thickness of the donor layer. The red and green shading of the boxes are for quick reference indicating quenching and enhancement, respectively. Single-photon non-resonant excitation is performed with 532 nm while two-photon non-resonant excitation is performed with 1040 nm.

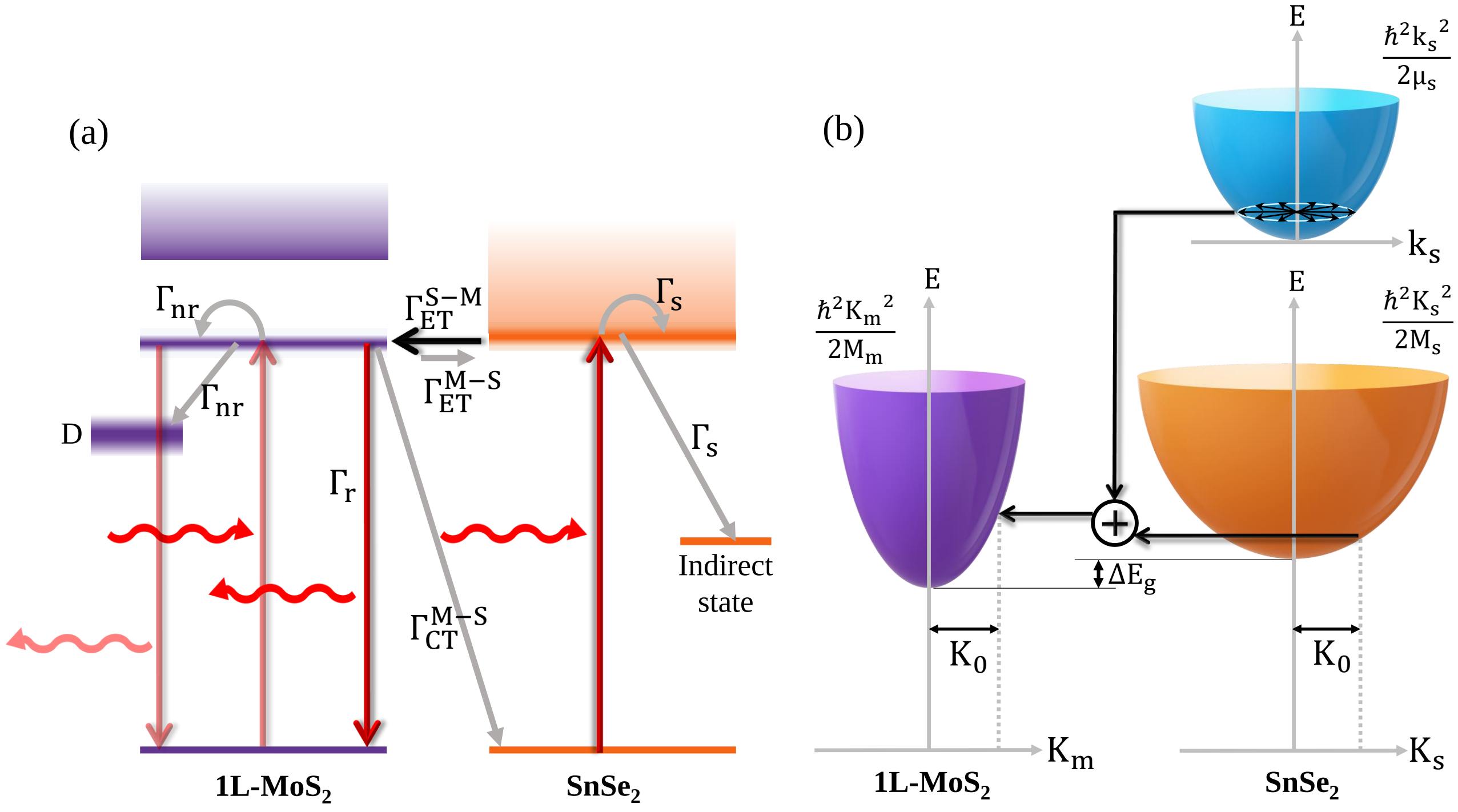

**Figure 4: FRET across 1L-$MoS_2$/$SnSe_2$.** (a) Transition diagrams of 1L-$MoS_2$ and $SnSe_2$ across the junction with 633 nm excitation. Corresponding processes following the excitations are labelled along the arrows indicating the direction of the process. (b) Two-particle band structure of excitons in $MoS_2$ (in violet, on the left) and free e-h pairs in $SnSe_2$ [including the center of mass component in orange (bottom right) and relative component in blue (top right)], indicating energy and center of mass momentum conservation during FRET process.

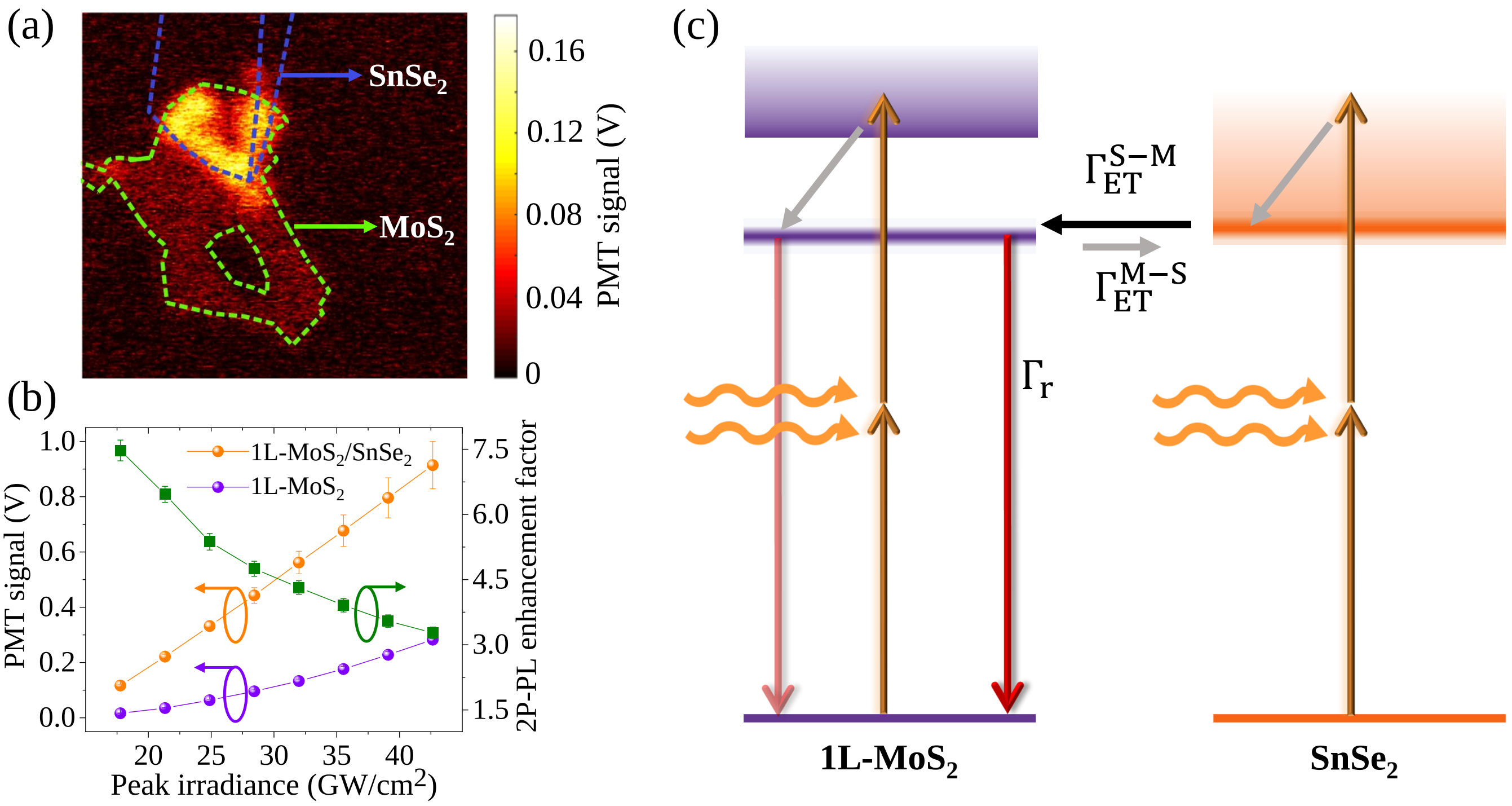

**Figure 5: Two-photon luminescence enhancement**. (a) Two-photon PL map of 1L-$MoS_2$(green) / $SnSe_2$(blue) of sample J1 of Fig.1(c) at 650 nm under two-photon excitation with 1040 nm laser at a peak irradiance of 17.8 GW/cm$^2$. Two-photon PL intensity corresponds to the voltage detected by Photo Multiplier Tube (PMT) whose range is represented by the color bar. (b) PMT detected voltage (left axis) of the two-photon PL of isolated 1L-$MoS_2$ and 1L-$MoS_2$/$SnSe_2$ junction, and the enhancement factor at the junction (right axis) as a function of peak irradiance of the excitation. (c) Transitions in 1L-$MoS_2$, $SnSe_2$ and across the junction with non-resonant two-photon excitation.

# Supporting Information

# Strong Single- and Two-Photon Luminescence Enhancement by Non-Radiative Energy Transfer across Layered Heterostructure

Medha Dandu, Rabindra Biswas, Sarthak Das, Sangeeth Kallatt, Suman Chatterjee, Mehak Mahajan, Varun Raghunathan, Kausik Majumdar*

Department of Electrical Communication Engineering, Indian Institute of Science, Bangalore 560012, India

*Corresponding author, email: *kausikm@iisc.ac.in*

**Different layered heterostructures used for PL measurements and Raman characterization**

Figure S1a shows the optical image of the 1L-$MoS_2$/$SnSe_2$ junction (sample J1) used for PL spectroscopy and mapping. 1L-$MoS_2$ is characterized with Raman spectroscopy with 532 nm for $A_{1g}$ and $E_{2g}^1$ modes which were obtained at 404.8 $cm^{-1}$ and 385.1 $cm^{-1}$ respectively.[1] Raman shift at the $SnSe_2$ region shows its $E_g$ and $A_{1g}$ peaks at 110.6 $cm^{-1}$ and 184.8 $cm^{-1}$ respectively. On the junction region, $A_{1g}$ and $E_{2g}^1$ modes of 1L-$MoS_2$ are quenched completely as shown in Figure S1b. Figure S1c and S1d show the optical images of 1L-$MoS_2$/$TaS_2$ and 1L-$WS_2$/$SnSe_2$ junctions used for PL characterization in this work.

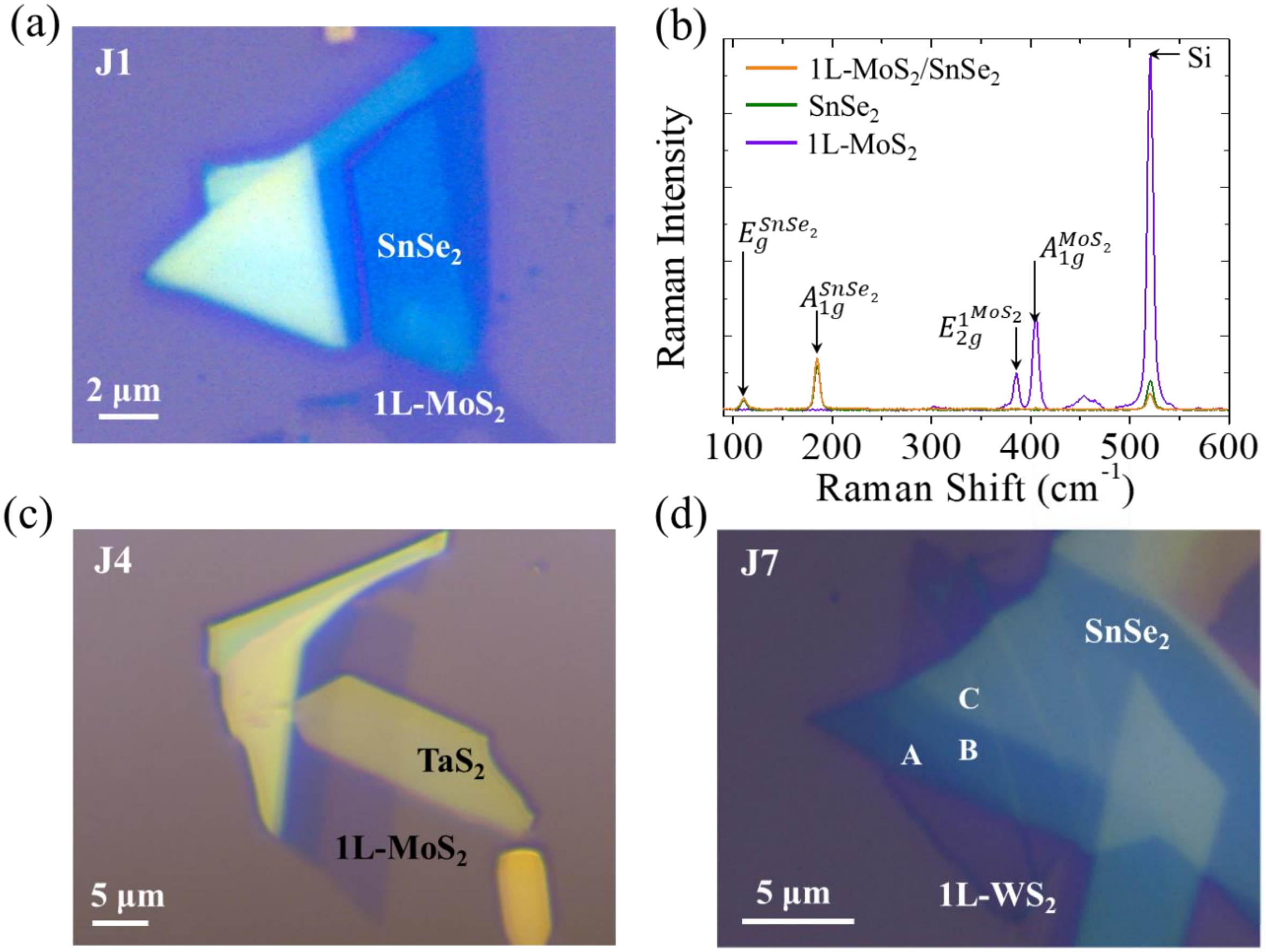

Figure S1: (a) Optical image of the 1L-$MoS_2$/$SnSe_2$ junction (J1) described in Figure 1a. (b) Raman shift of 1L-$MoS_2$, $SnSe_2$ and the junction from sample J1 characterized with 532 nm laser. (c) Optical image of the 1L-$MoS_2$/$TaS_2$ junction J4 whose spectra are represented in Figure 1f. (d) Optical image of the 1L-$WS_2$/$SnSe_2$ junction J7 described in figure 2e-f.

**PL enhancement across 1L-$MoS_2$/$SnSe_2$ from multiple samples (J2 and J3)**

PL enhancement of 1L-$MoS_2$ on 1L-$MoS_2$/$SnSe_2$ junction is verified across other samples J2 and J3 with both 633 nm and 532 nm excitation.

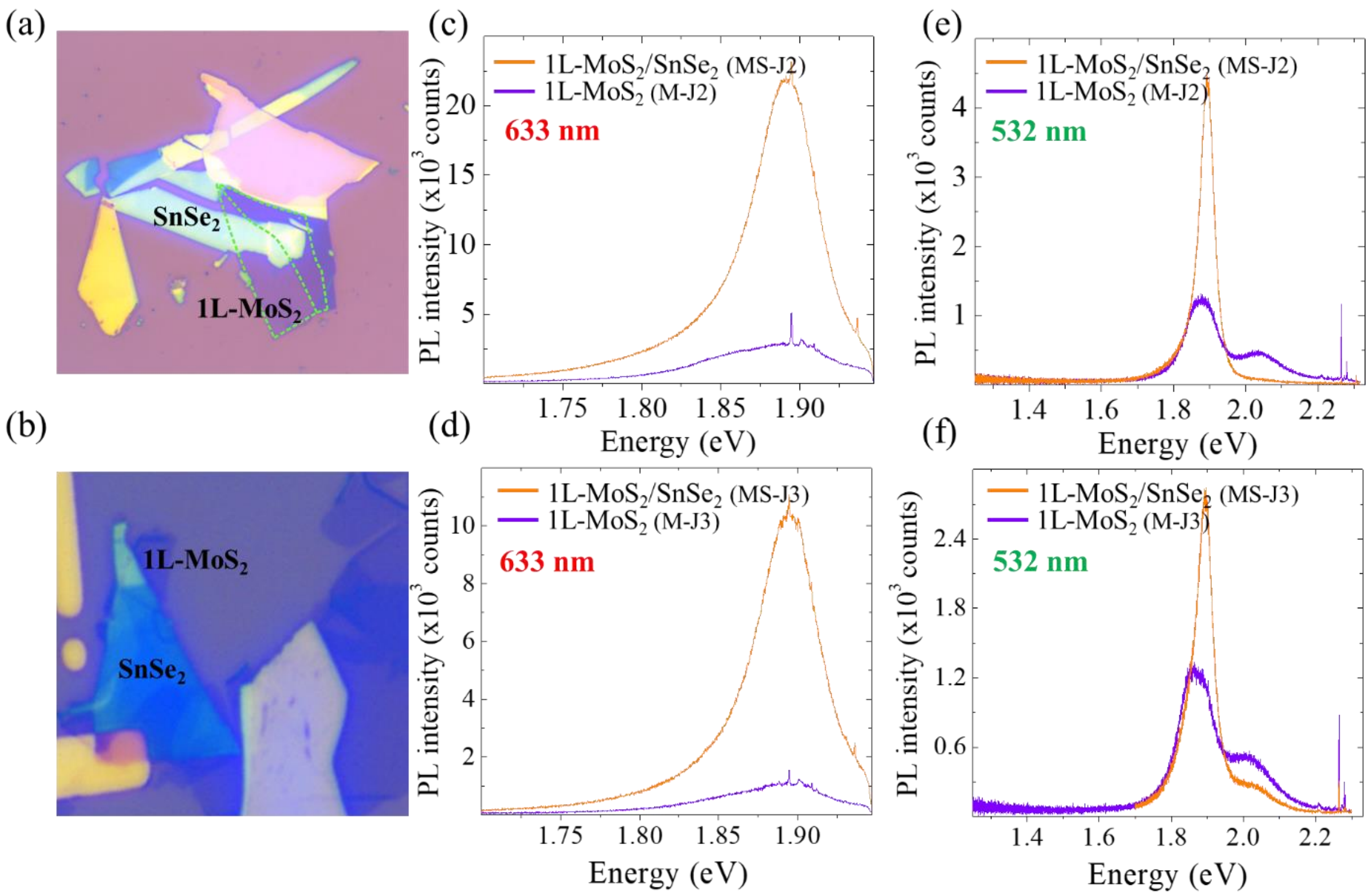

Figure S2: (a-b) The optical images of 1L-$MoS_2$/$SnSe_2$ junctions from samples J2 and J3. Sample J2 is used to characterize temperature dependence of PL enhancement. (c-d) 633 nm PL spectra of isolated 1L-$MoS_2$ (violet) and junction (orange) regions. (e-f) 532 nm PL spectra of isolated 1L-$MoS_2$ (violet) and junction (orange) regions.

**Photoluminescence spectroscopy across another 1L-$MoS_2$/hBN/$SnSe_2$ sample (J6)**

Due to the manifestation of efficient FRET across 1L-$MoS_2$/$SnSe_2$, PL enhancement persists even with insertion of spacer layer, hBN as discussed in the main text from sample J5. We verify this PL enhancement with both 633 nm and 532 nm across 1L-$MoS_2$/hBN/$SnSe_2$ with another sample J6.

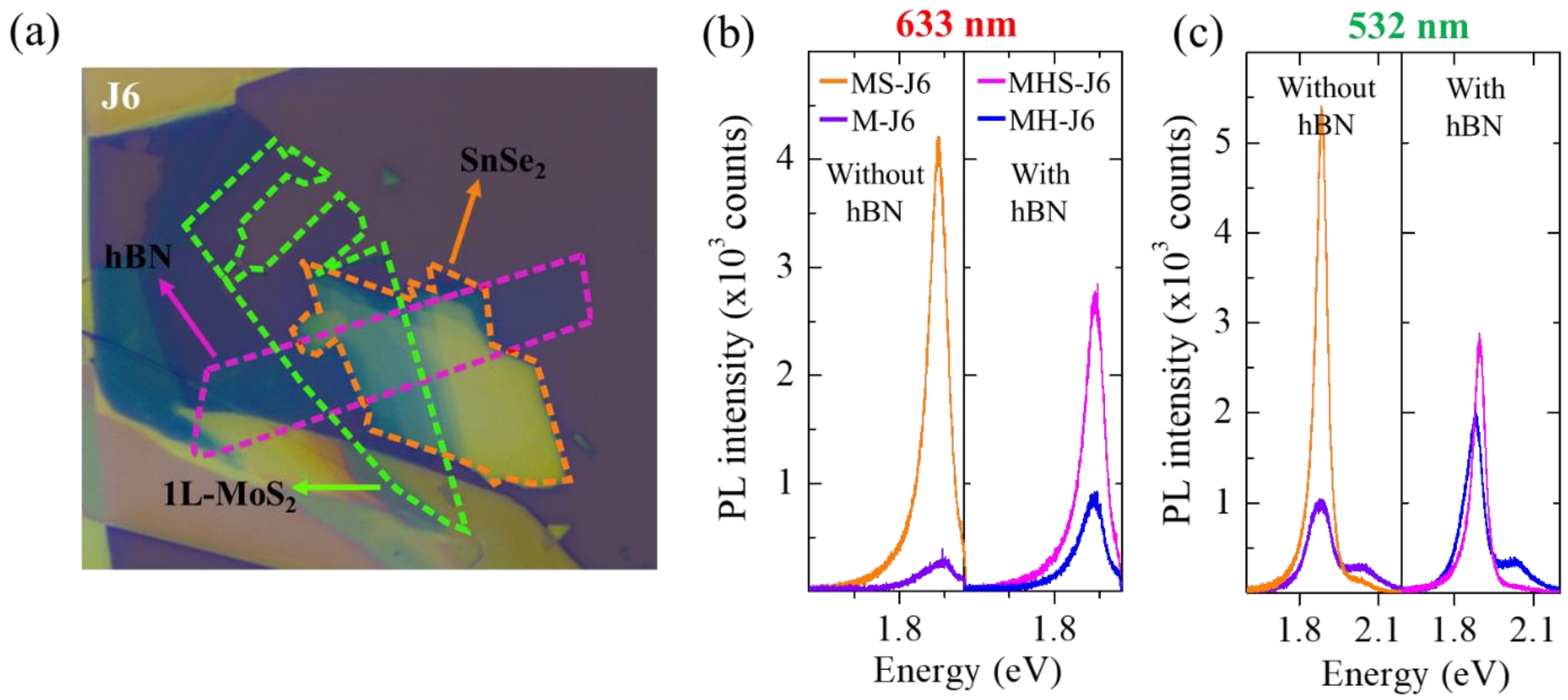

Figure S3: (a) Optical image of sample J6 used to verify PL enhancement of 1L-$MoS_2$ on the 1L-$MoS_2$/hBN/$SnSe_2$ junction. (b-c) 633 nm and 532 nm PL spectra of 1L-$MoS_2$/$SnSe_2$ junction and the corresponding isolated 1L-$MoS_2$ control without (left panel) and with the presence of hBN (right panel) in the sample J6.

**Reflectance contrast of $SnSe_2$ at low temperature**

Reflectance contrast spectra of $SnSe_2$ flakes of different thickness on $SiO_2$/Si substrate are also measured at 4K. The dip in the reflectance contrast is found to change its spectral position with $SnSe_2$ thickness as observed at room temperature. Even at this low temperature, the dip is still spectrally broad and does not show any sharp excitonic feature. This helps to confirm that transition dipole in $SnSe_2$ is governed by free electron-hole pairs as the screening offered by the large carrier concentration suppresses strongly bound exciton state in $SnSe_2$.

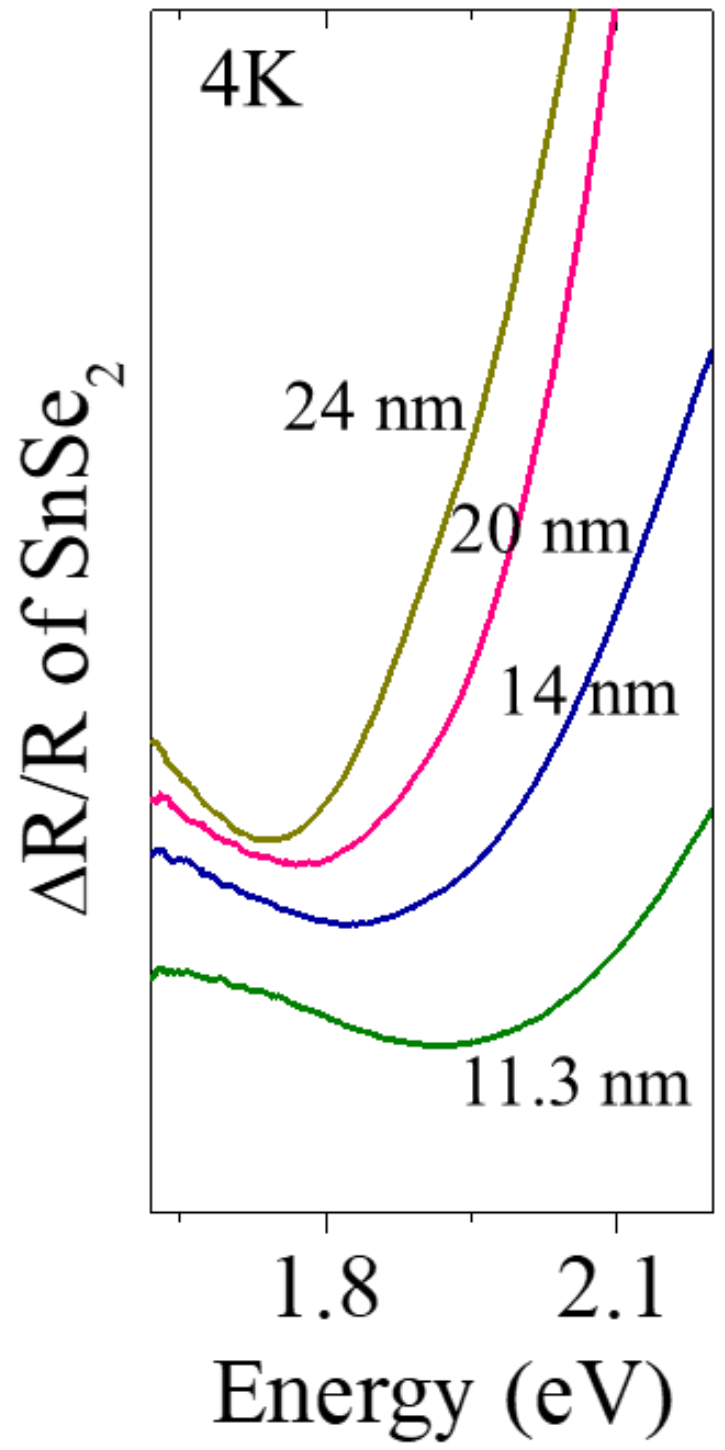

Figure S4: Reflectance contrast spectra of isolated $SnSe_2$ flakes on $SiO_2$/Si substrate at 4 K. Corresponding thickness of $SnSe_2$ is indicated at each spectrum.

**Photoluminescence spectroscopy across 2L-$MoS_2$/$SnSe_2$**

Unlike 1L-$MoS_2$, 2L-$MoS_2$ does not exhibit any PL enhancement on the junction with $SnSe_2$. Figure S5a shows the quenched PL spectra of 2L-$MoS_2$/$SnSe_2$ with 633 nm at different temperatures. From Figure S5b we can see that there is decrease in PL quenching with increase in temperature like that of 1L-$MoS_2$/$SnSe_2$. The absence of PL enhancement in the case of 2L-$MoS_2$/$SnSe_2$ can be correlated with the competition between the rate of FRET and increased non-radiative scattering in bilayer as seen from the model in the next section.

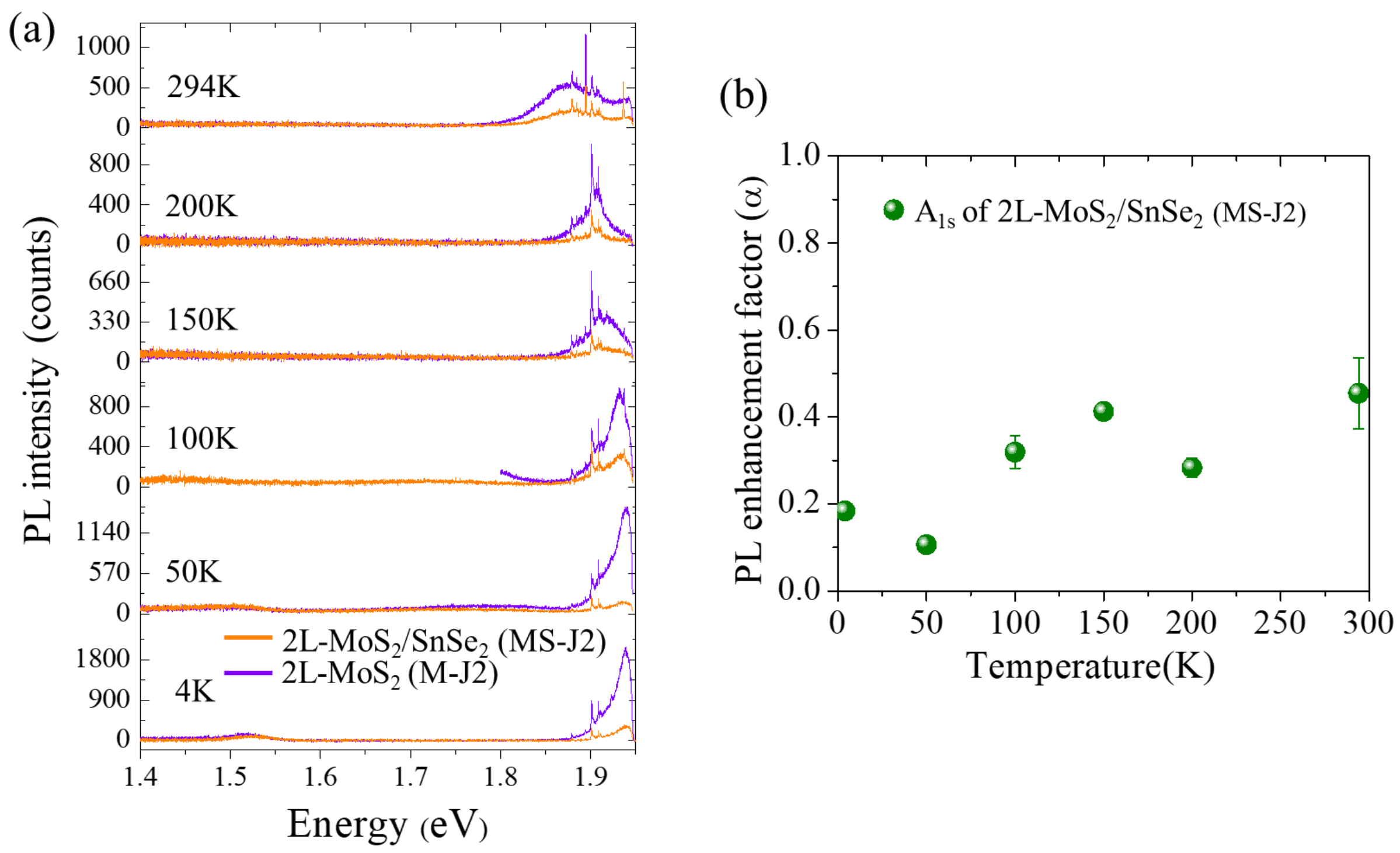

Figure S5: (a) Temperature dependent PL spectra of 2L-$MoS_2$ (violet) and its junction with $SnSe_2$ (orange) from sample J2 under 633 nm excitation. (b) Variation of PL enhancement factor (α) of $A_{1s}$ exciton peak of 2L-$MoS_2$/$SnSe_2$ with 633 nm excitation.

**Discussion on the rate of FRET and PL enhancement factor**

**Deriving PL enhancement factor from the rate equations:**

Experimental data proves that the PL enhancement observed across different layered heterostructures is sensitive to excitation wavelength (resonant versus non-resonant), temperature, donor and acceptor materials, thickness of the donor and the acceptor, and the separation between them. To get a qualitative understanding of the effect of these parameters, we derive an expression for the PL enhancement factor from the rate equations of exciton density in the isolated 1L-$MoS_2$ and the 1L-$MoS_2$ /$SnSe_2$ junction.

Rate equations for Monolayer $MoS_2$ on $SiO_2$ with resonant excitation:

633 nm excitation near resonantly excites the exciton states of 1L-$MoS_2$. Let $G_{ex}^{M}$ be the generation rate of $A_{1s}$ excitons with 633 nm in 1L-$MoS_2$ on $SiO_2$. These excitons can either recombine radiatively at a rate $\Gamma_r$ to give out photoluminescence or decay non-radiatively at a rate $\Gamma_{nr}^{M}$. The non-radiative decay ($\Gamma_{nr}$) comprises of scattering to the defect states at lower energies and scattering to the energy states outside the light cone as schematically shown in Figure 4b. The rate equation for the exciton density ($N_{ex}^{M}$) density in isolated 1L-$MoS_2$ on $SiO_2$ can be written as the following.

$$\frac{dN_{ex}^{M}}{dt} = G_{ex}^{M} - (\Gamma_r + \Gamma_{nr}^{M})N_{ex}^{M} \tag{1}$$

The steady state photoluminescence signal from a given excitation will be proportional to the fraction of steady state exciton density, $N_{ex}^{M} = \frac{G_{ex}^{M}}{(\Gamma_r+\Gamma_{nr}^{M})}$ decaying radiatively. So, the PL intensity from 1L-$MoS_2$ on $SiO_2$ is given to be

$$I_M \propto \frac{G_{ex}^{M}\Gamma_r}{(\Gamma_r + \Gamma_{nr}^{M})} \tag{2}$$

Rate equations for Monolayer $MoS_2$ on $SnSe_2$ with resonant excitation:

As shown in the schematic, 633 nm near resonantly excites multi-layer $SnSe_2$ along with 1L-$MoS_2$. This excitation creates free electron hole pairs in $SnSe_2$ at its direct bandgap at a rate $G_{e-h}^{S}$ (not excitons because of screening of carriers in the degenerately doped $SnSe_2$). These free electron hole pairs are coupled to the $A_{1s}$ exciton states via FRET at a rate of $\Gamma_{ET}^{S-M}$ or they can be scattered from the energy states in $SnSe_2$ (resonant to $A_{1s}$ of $MoS_2$) to other non-resonant energy states predominantly to the states at the indirect band gap at a rate $\Gamma_s$. Excitons in 1L-$MoS_2$ either excited with 633 nm on the junction or created with FRET from $SnSe_2$ will also be coupled to the states in $SnSe_2$ through the process of FRET and thus they can decay to $SnSe_2$ by this dipole-dipole coupling at a rate denoted by $\Gamma_{ET}^{M-S}$. Electrons and holes from exciton state in $MoS_2$ can also dissociate to lower energy states in conduction and valence bands of $SnSe_2$ which causes decay of excitons through charge transfer at a rate of $\Gamma_{CT}^{M-S}$. The non-radiative decay rate of excitons on the junction $\Gamma_{nr}^{jun}$ can be different from $\Gamma_{nr}^{M}$ as seen from the change of broadening of PL peaks on the junction. However, the radiative decay rate $\Gamma_r$ is assumed to similar in 1L-MoS2 and the junction. Generation rate of free e-h pairs in $SnSe_2$ and excitons in $MoS_2$ on the junction are assumed to be similar to the rates in individual layers. From the above discussion, the rate equations for exciton density in 1L-$MoS_2$ on the junction ($N_{ex}^{M,jun}$) and free e-h pair density in $SnSe_2$ ($N_{e-h}^{S,jun}$) can be expressed as

$$\frac{dN_{ex}^{M,jun}}{dt} = G_{ex}^{M} - \left(\Gamma_r + \Gamma_{nr}^{M,jun} + \Gamma_{CT}^{M-S} + \Gamma_{ET}^{M-S}\right)N_{ex}^{M,jun} + \Gamma_{ET}^{S-M}N_{e-h}^{S,jun} \qquad (3)$$

$$\frac{dN_{e-h}^{S,jun}}{dt} = G_{e-h}^{S} - \left(\Gamma_s + \Gamma_{ET}^{S-M}\right)N_{e-h}^{S,jun} + \left(\Gamma_{CT}^{M-S} + \Gamma_{ET}^{M-S}\right)N_{ex}^{M,jun} \qquad (4)$$

Since an order of enhancement is seen in the PL intensity of 1L-$MoS_2$ on the junction, the contribution of $G_{ex}^{M}$ in (3) can be neglected. From (3) and (4), $I_{M,jun}$ can be obtained from $N_{ex}^{M,jun}$ as

$$N_{ex}^{M,jun} = \frac{G_{e-h}^{S}\Gamma_{ET}^{S-M}}{\Gamma_s\left(\Gamma_r + \Gamma_{nr}^{M,jun} + \Gamma_{CT}^{M-S} + \Gamma_{ET}^{M-S}\right) + \Gamma_{ET}^{S-M}\left(\Gamma_r + \Gamma_{nr}^{M,jun}\right)} \qquad (5)$$

$$I_{M,jun} \propto N_{ex}^{M,jun}\Gamma_r \qquad (6)$$

The PL enhancement factor, α is the ratio of steady state PL intensity of $A_{1s}$ peak from 1L-$MoS_2$/$SnSe_2$ ($I_{M,jun}$) to that of 1L-$MoS_2$ on $SiO_2$ ($I_M$).

$$\alpha = \frac{\left(\frac{G_{e-h}^{S}}{G_{ex}^{M}}\right)\Gamma_{ET}^{S-M}}{\Gamma_s\left(\frac{\Gamma_r + \Gamma_{nr}^{M,jun}}{\Gamma_r + \Gamma_{nr}^{M}} + \frac{\Gamma_{CT}^{M-S} + \Gamma_{ET}^{M-S}}{\Gamma_r + \Gamma_{nr}^{M}}\right) + \Gamma_{ET}^{S-M}\left(\frac{\Gamma_r + \Gamma_{nr}^{M,jun}}{\Gamma_r + \Gamma_{nr}^{M}}\right)} \quad (7)$$

To get a preliminary insight into the role of different rate parameters on α, it is reasonable to ignore the difference between $\Gamma_{nr}^{M,jun}$ and $\Gamma_{nr}^{M}$. α can now be expressed as

$$\alpha = \left(\frac{G_{e-h}^{S}}{G_{ex}^{M}}\right)\frac{\gamma}{\gamma + \beta} \quad (8)$$

where $\gamma = \frac{\Gamma_{ET}^{S-M}}{\Gamma_s}$ and $\beta = 1 + \left(\frac{\Gamma_{CT}^{M-S}+\Gamma_{ET}^{M-S}}{\Gamma_r+\Gamma_{nr}^{M}}\right)$.

**Temperature dependence of the rate of FRET:**

As seen from eq-(7), variation of α with temperature can be related to the temperature dependent rate of different processes on illumination. $\Gamma_{CT}$ and $\Gamma_r$ are very weakly dependent on the temperature while $\Gamma_{ET}$ and $\Gamma_{nr}$ that vary with temperature can play a role in determining the value of α. We adopt the energy transfer model by Lyo[2] based on dipole-dipole coupling across quantum wells to understand the temperature dependence of α in our case. This model deals with the energy transfer from an exciton to a free e-h pair which predicts its rate as

$$\Gamma_{ET} = \Gamma_{ET}^{0}(\xi_T)g\left(\frac{d}{\xi_T}\right) \quad (9)$$

where d is the physical separation between the donor and the acceptor layers and $\xi_T$ is a temperature dependent parameter, $\xi_T = \sqrt{\frac{\hbar^2}{2Mk_BT}}$ where $k_B$ is the Boltzmann constant and M is the total mass of exciton ($M=m_e+m_h$) which is assumed to be same for both the layers.

$\Gamma_{ET}^{0}(\xi_T) = \frac{32\pi\mu}{\hbar^3}\left(\frac{q^2D_1D_2}{\kappa a_B\xi_T}\right)^2$ which depends the system of donor and acceptor with $\mu$ as reduced exciton mass, $a_B$ is the exciton bohr radius, $\kappa$ is the average dielectric constant, $D_1$ and $D_2$ are the dipole moments of the donor and the acceptor. g(t) is expressed as

$$g(t) = \int_0^\infty x^3 e^{-x^2} e^{-2tx} S(tx)^2\, dx \quad (10)$$

with $s(t) = \frac{\sin\left(\frac{tb_1}{2d}\right)}{\frac{tb_1}{2d}} \frac{\sin\left(\frac{tb_2}{2d}\right)}{\frac{tb_2}{2d}}$ , where $b_1$ and $b_2$ are the thickness of the donor and the acceptor respectively. To project the variation of FRET in our case, we assume that rate of energy transfer is of similar magnitude in both the directions across $MoS_2$ and $SnSe_2$ and therefore (9) can be applicable to calculate both $\Gamma_{ET}^{S-M}$ and $\Gamma_{ET}^{M-S}$. Assuming $b_1$=10 nm ($SnSe_2$), $b_2$=1 nm ($MoS_2$), d=1 nm and M ~ $m_o$ in $SnSe_2$,[3] we obtain the temperature dependence of $\Gamma_{ET}$ as shown in Figure S6a.

Using the above variation of $\Gamma_{ET}$ and (8) with assumption of parameters $\Gamma_{CT} = 10^{12}$ $s^{-1}$, $\Gamma_{nr} + \Gamma_r = 0.5\text{x}10^{12}$ $s^{-1}$, $\Gamma_s = 10^{13}$ $s^{-1}$ and $\left(\frac{G_{e-h}^S}{G_{ex}^M}\right)$=10, an evaluation of α is done at multiple temperature which matches with the trend of experimental data as in Figure S6b.

(8) also explains the absence of enhancement for trion in 1L-$MoS_2$ from the interplay between FRET and radiative recombination. As Figure S6c shows the variation of α with $\Gamma_{ET}$ and $\Gamma_r$ keeping all other rate parameters fixed. $\alpha > 1$ for those cases where $\beta < \gamma$ which indicates that $\Gamma_r$ should be fast enough to yield luminescence of FRET coupled dipoles in the acceptor. It has been shown experimentally that trion has a longer life time than exciton which makes $\Gamma_r$ of exciton higher than that of trion.[4] This increased lifetime of trion accelerates their loss from $MoS_2$ to $SnSe_2$ and hinders the luminescence of trions from energy transferred dipoles.

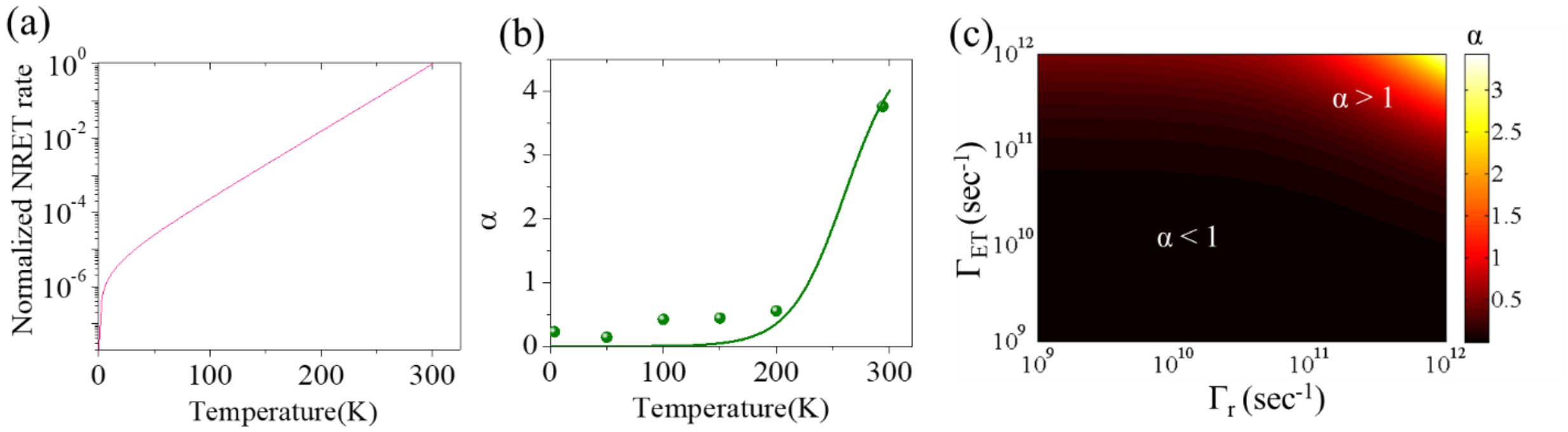

Figure S6: (a) Variation of the rate of FRET with temperature calculated from theoretical model. The y-axis is normalized with respect to the FRET rate at 300K. (b) Experimental (symbol) and model predicted (line) values of enhancement factor (α) calculated at different temperatures with 532 nm excitation. (c) Color plot of variation of enhancement factor (α) with change in the rate of FRET and the rate of radiative recombination. The enhancement ($\alpha > 1$) and quenching ($\alpha < 1$) regions are indicated.